\documentclass[apj]{emulateapj}
\usepackage{graphicx}
\usepackage{amssymb}
\usepackage{amsmath}
\usepackage{epstopdf}

\shorttitle{NIR Proper Motions of Magnetars}
\shortauthors{Tendulkar, Cameron \& Kulkarni}

\begin{document}
\setlength{\fboxsep}{0pt}%
\setlength{\fboxrule}{0.5pt}%

\title{Proper Motions and Origins of SGR\,1806$-$20 and SGR\,1900$+$14}
\author{Shriharsh P. Tendulkar}
\affil{California Institute of Technology}
\affil{1200 E California Blvd, MC 249-17, Pasadena, CA 91125, USA}
\email{spt@astro.caltech.edu}

\author{P. Brian Cameron}
\affil{The Aerospace Corporation}
\affil{15049 Conference Center Drive, Chantilly, VA 20151-3824, USA}
\email{pbc@astro.caltech.edu}

\author{Shrinivas R. Kulkarni}
\affil{California Institute of Technology}
\affil{1200 E California Blvd, MC 249-17, Pasadena, CA 91125, USA}
\email{srk@astro.caltech.edu}


\begin{abstract}
We present results from high-resolution infrared observations of magnetars SGR\,1806$-$20 and SGR\,1900$+$14 over 5 years using laser-supported adaptive optics at the 10-m Keck Observatory. Our measurements of the proper motions of these magnetars provide robust links between magnetars and their progenitors and provide age estimates for magnetars. At the measured distances of their putative associations, we measure the linear transverse velocity of SGR\,1806$-$20 to be $350\pm100\,\mathrm{km\,s^{-1}}$ and of SGR\,1900$+$14 to be $130\pm30\,\mathrm{km\,s^{-1}}$. The transverse velocity vectors for both magnetars point away from the clusters of massive stars, solidifying their proposed associations. Assuming that the magnetars were born in the clusters, we can estimate the braking index to be  $\sim 1.8$ for SGR\,1806$-$20 and $\sim 1.2$ for SGR\,1900$+$14. This is significantly lower than the canonical value of $n=3$ predicted by the magnetic dipole spin-down suggesting an alternative source of dissipation such as twisted magnetospheres or particle winds.
\end{abstract}

\maketitle

\section{Introduction}
\label{sec:introduction}
Magnetars were proposed~\citep{thompson1995, thompson1996} as a unified model to explain the phenomena of soft gamma repeaters (SGRs) and anomalous X-ray pulsars (AXPs). Magnetars, unlike canonical radio pulsars, would have a very high magnetic field strength $B$ ($\sim 10^{14}\,\mathrm{G}$) such that their internal energy was dominated by their magnetic energy rather than their rotational energy. The SGR flares were explained as resulting from violent magnetic reconnections and crustal quakes and the quiescent X-ray emission of AXPs (which is much larger than their spin-down luminosity) was attributed to the decay of intense magnetic fields. The discovery of large period derivatives \citep[$\dot{P}$ $\sim 10^{-10}\,\mathrm{s\,s^{-1}}$;][]{kouveliotou1998} confirmed the basic expectation of the magnetar model. For recent reviews of observational and theoretical progress in the field we refer the readers to ~\citet{mereghetti2008} and \citet{hurley2011}.

Despite the successes of the magnetar model, we have little understanding of why only some neutron stars are born as magnetars. Originally, \citet{thompson1993} invoked a rapidly spinning ($\sim$ one to three millisecond) proto-neutron star as essential for strong amplification of a seed magnetic field. The rapidly spinning neutron stars would result in a supernova more energetic than a canonical core-collapse supernova.

The observational support for the formation mechanism of magnetars appears to be lacking. \citet{vink2006} showed that the three supernova remnants (SNRs) to which three magnetars are best paired, (Kes\,73, CTB\,109 and N49), are completely consistent with the standard supernova explosion energies.

The offset between SGR\,0525$-$66 (previously known as ``5 March 1979'') and its surrounding supernova remnant N49 and the notion that some halo SGRs might explain a fraction of GRBs led to the expectation of SGRs having high space motion \citep[see ][]{rl96}. This spawned a number of efforts  to measure the space motions of magnetars.

Here, we present astrometric observations of two of the youngest magnetars: SGR\,1806$-$20 and SGR\,1900+14. The resulting measurements of proper motion allow us to trace back these two objects to their potential birth sites and additionally measure the space motions as well.  The paper is organized as follows. In Section~\ref{sec:targets}, we summarize our knowledge of these two magnetars. In Section~\ref{sec:observations}, we describe our observations, data reduction methodology and analysis techniques for point spread function (PSF) fitting, relative astrometry and photometry. We present the results in Sections~\ref{sec:results} and in Section~\ref{sec:discussion} we discuss the significance of our proper motion measurements. 

\section{Targets}
\label{sec:targets}

\begin{deluxetable}{lcc}
\centering
\tablecolumns{3} 
\tablecaption{Characteristics of SGR\,1806$-$20 and SGR\,1900$+$14.\label{tab:sgr_character}}
\tablewidth{0pt}
\tabletypesize{\footnotesize}
\tablehead{
  \colhead{}                  &
  \colhead{SGR\,1806$-$20}       &
  \colhead{SGR\,1900$+$14}               \\
}
\startdata
Period\,$P\,\mathrm{(sec)}$ & 7.6022(7)  & 5.19987(7)\\
$\dot{P}\,\mathrm{(10^{-11}\,s\,s^{-1})}$ \tablenotemark{a} & 49 &  17 \\
$P/\dot{P}\,\mathrm{(kyr)}$ & 0.32 & 1.8\\
$B_{\mathrm{Surf}}\,\mathrm{(10^{14}\,G)}$ & 24  & 7.0 \\
R.A (J2000) & $18^{\mathrm{h}}\,08^{\mathrm{m}}\,39.337^{\mathrm{s}}$  & $19^{\mathrm{h}}\,07^{\mathrm{m}}\,14.31^{\mathrm{s}}$\\
Dec (J2000) & $−20^{\circ}\,24^{\prime}\,39.85^{\prime\prime}$  &  $9^{\circ}\,19^{\prime}\,19.74^{\prime\prime}$\\
\enddata
\tablenotetext{a}{Average period derivative calculated from X-ray period measurements from literature. See Section~\ref{sec:braking_index}}
\tablecomments{Refer to http://www.physics.mcgill.ca/$\sim$pulsar/\\magnetar/main.html. Positions are from \textit{Chandra} X-ray observations. }
\end{deluxetable}

Table~\ref{tab:sgr_character} summarizes the essential characteristics of both our targets; SGR\,1806$-$20 and SGR\,1900$+$14. We discuss each target in further detail in the following sections.

\subsection{SGR\,1806$-$20}
SGR\,1806$-$20 (previously known as GB790107) was identified as a repeating gamma-ray burst with a soft spectrum by~\citet{laros1986}.
SGR\,1806$-$20 is best known for its giant burst of December 27, 2004~\citep{hurley2005,palmer2005} which was one of the brightest cosmic flares ever detected. The burst was followed by a long lived radio afterglow~\citep{cameron2005,gaensler2005,spreeuw2010} which allowed the precise localization of the source.

\subsubsection{Association with Star Cluster}
SGR\,1806$-$20 lies in a radio nebula G10.0-0.3~\citep{kulkarni1995} which is a part of the W31 $\mathrm{H II}$ complex. It was earlier suggested that the massive star LBV 1806$-$20 and its surrounding radio nebula were associated with SGR\,1806$-$20~\citep{vankerkwijk1995} but precise \textit{Chandra} localization~\citep{kaplan2002a} proved that SGR\,1806$-$20 was 14$^{\prime\prime}$ away from the center of G10.0-0.3 and 12$^{\prime\prime}$ away from LBV 1806$-$20. A cluster of massive stars, coincident with a mid-IR nebulosity, was discovered by \citet{fuchs1999} about $7^{\prime\prime}$ to the north of the magnetar. 

\begin{deluxetable*}{llcl}
\centering
\tablecaption{Distance to SGR\,1806$-$20 measured by various authors.\label{tab:sgr1806_distance}}
\tablewidth{0pt}
\tabletypesize{\footnotesize}
\tablehead{
  \colhead{} &
  \colhead{Reference} &
  \colhead{Distance} &
  \colhead{Comments} \\
  \colhead{} &
  \colhead{} &
  \colhead{(kpc)} &
  \colhead{} \\
}
\startdata

1 & \citet{cameron2005} & $6.5-9.8$ & HI absorption from Dec 2004 flare \\
2 & \citet{mcclure2005} & $> 6.5$ & HI absorption from Dec 2004 flare \\
3 & \citet{svirski2011} & $9.4 - 18.6$ & X-ray scattering echos \\
4 & \citet{bibby2008} & $8.7^{+1.8}_{-1.5}$ & Spectral classification, IR photometry\\
  &                   &                    & and cluster isochrones \\
5 & \citet{figer2004} & $11.8 \pm 0.5$ & Radial Velocity (RV) of LBV 1806 \\
6 &\citet{eikenberry2004} & $15^{+1.8}_{-1.3}$ & RV of LBV 1806 and surrounding nebula\\
  &                       &                   &  and Galactic rotation curve  \\
  &                       & strictly $> 9.5$ & Luminosity of cluster stars \\
  &                       & strictly $>5.7$ & Ammonia absorption to LBV 1806 \\

\enddata
\end{deluxetable*}

Table~\ref{tab:sgr1806_distance} lists all the distance measurements reported to date. We place a higher premium for distance estimates related to the X-ray counterpart of SGR\,1806$-$20 or the associated cluster of massive stars over the estimates to LBV\,1806$-$20, since it is unclear whether LBV\,1806$-$20 is physically near the magnetar. In Table~\ref{tab:sgr1806_distance}, measurements 1--4 are distances to SGR\,1806$-$20 or the cluster of massive stars and measurements 5 and 6 are distances to LBV\,1806$-20$. We adopt a nominal distance of $9 \pm 2\,\mathrm{kpc}$ which is consistent with all the measurements.

\subsubsection{IR Counterpart}

\begin{figure}[htb]
\centering
\fbox{\includegraphics[width=0.48\textwidth]{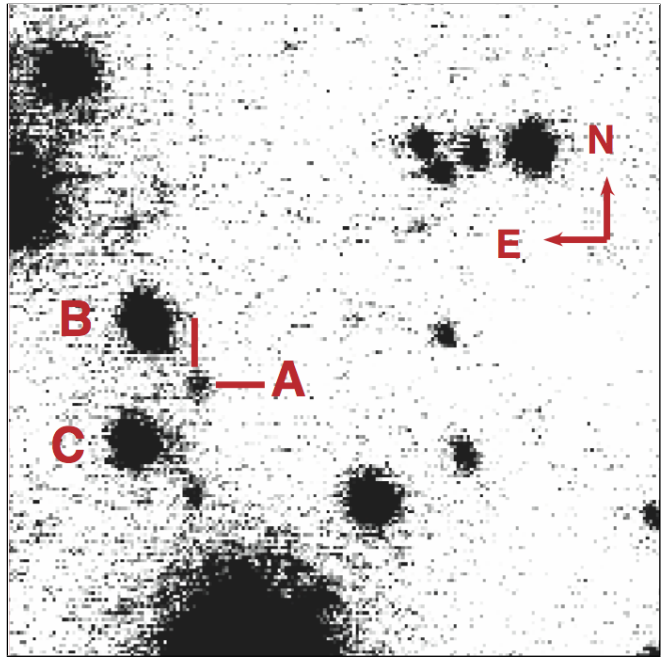}}
\caption{A $2\times2\,\mathrm{arcsec}$ cutout near SGR\,1806$-$20 from a $\mathrm{K_s}$ band LGS-AO supported observation from the NIRC2 camera. The IR counterpart, as identified by~\citep{kosugi2005,israel2005} is marked with cross hairs and labeled A as per \citep{israel2005} as are stars B and C.}
\label{fig:sgr1806_IR_closeup}
\end{figure}

Figure~\ref{fig:sgr1806_IR_closeup} shows a $2\times2\,\mathrm{arcsec}$ cutout near SGR\,1806$-$20 from our laser guide star adaptive optics (LGS-AO) supported observations in the $\mathrm{K_s}$ band using the NIRC2 instrument (See Section~\ref{sec:observations} for details). Star A was suggested as the NIR counterpart for SGR\,1806$-$20 by ~\citet{kosugi2005} and independently by~\citet{israel2005} based on NIR variability over the 2004 active period. Using the NAOS-CONICA instrument on the 8.1-m Very Large Telescope, \citet{israel2005} monitored SGR\,1806$-$20 on 11 epochs between March and October 2004. They measured a factor of two increase in the flux of the star A with a $>$ 9-$\sigma$ confidence. The IR flux increase corresponded well with X-ray flux that also increased by a factor of two in the 2--10\,keV and  20--100\,keV bands~\citep[\textit{XMM-Newton, INTEGRAL};][]{mereghetti2005b,mereghetti2005a}. Our photometric measurements show a factor of three variability in the brightness of the same object (Section~\ref{sec:sgr1806_results}). The identification of the IR counterpart of SGR\,1806$-$20 appears to be secure.

\subsection{SGR\,1900$+$14}
The first bursts from SGR\,1900$+$14 (originally known as B1900$+$14) were identified by~\citet{mazets1979}. A very bright flare was detected on August 27 1998 with a  $\gamma$-ray peak followed by a 300-s long tail~\citep{hurley1999a,kouveliotou1999}. Following the burst, a fading radio~\citep{frail1999} and X-ray source~\citep{hurley1999a} was discovered. These observations led to a precise localization to within 0.15$^{\prime\prime}$.

\subsubsection{Association with Star Cluster}
SGR\,1900$+$14 is located near two objects from which it could have originated. A cluster of massive stars~\citep{vrba2000}, hidden behind two bright M5 super-giants, lies 12$^{\prime\prime}$ to the east of SGR\,1900$+$14 and a $10^4~\mathrm{yr}$ old, 12$^{\prime}$ diameter SNR G042.8+00.6 lies 17$^{\prime}$ to the south-east~\citep{mazets1979,kouveliotou1993,vasisht1994}. If SGR\,1900$+$14 was associated with the cluster of massive stars then it implies a young age and a space velocity close to the canonical value for pulsars. However, if it is associated with the supernova remnant then it would have a very high proper motion. An upper limit to the proper motion (based on \textit{Chandra} X-ray observatory imaging observations) of $\leq 100\,$milli-arcsecond\,yr$^{-1}$ is nominally inconsistent with the association of SGR\,1900$+$14 with the SNR ~\citep{kaplan2009,deluca2009}.

\citet{wachter2008} reported the discovery of an infrared elliptical ring or shell surrounding SGR\,1900$+$14 which was interpreted as a dust-free cavity created by the giant flare of August 1998. The authors concluded that SGR\,1900$+$14 is unambiguously associated with the afore mentioned star cluster. 

With adaptive-optics assisted Keck/NIRC2 imaging and Keck/NIRSPEC spectroscopy of the cluster near SGR\,1900,~\citet{davies2009} estimated the progenitor mass to be $17\pm2~\mathrm{M_{\odot}}$ which is much lower than the progenitor masses estimated for other magnetars ($\sim40$ to $50\,\mathrm{M_{\odot}}$).

\subsubsection{Distance}
\citet{vrba1996} showed that the bright IR sources noted by \citet{hartmann1996} at the \textit{ROSAT} localization of SGR\,1900$+$14 were M5 super-giant stars at a distance of 12 to 15\,kpc with an extinction of $A_V \approx19.2\,\mathrm{mag}$.  \citet{davies2009} measured a radial velocity of $-15.5\pm4\,\mathrm{km\,s^{-1}}$ for the cluster of stars implying a distance of $12.5 \pm 1.7\,\mathrm{kpc}$ using the measured model of Galactic rotation. We adopt the measurement of \citet{davies2009} for the distance to SGR\,1900$+$14.

\subsubsection{IR counterpart}

\begin{figure}[htb]
\centering
\fbox{\includegraphics[width=0.48\textwidth]{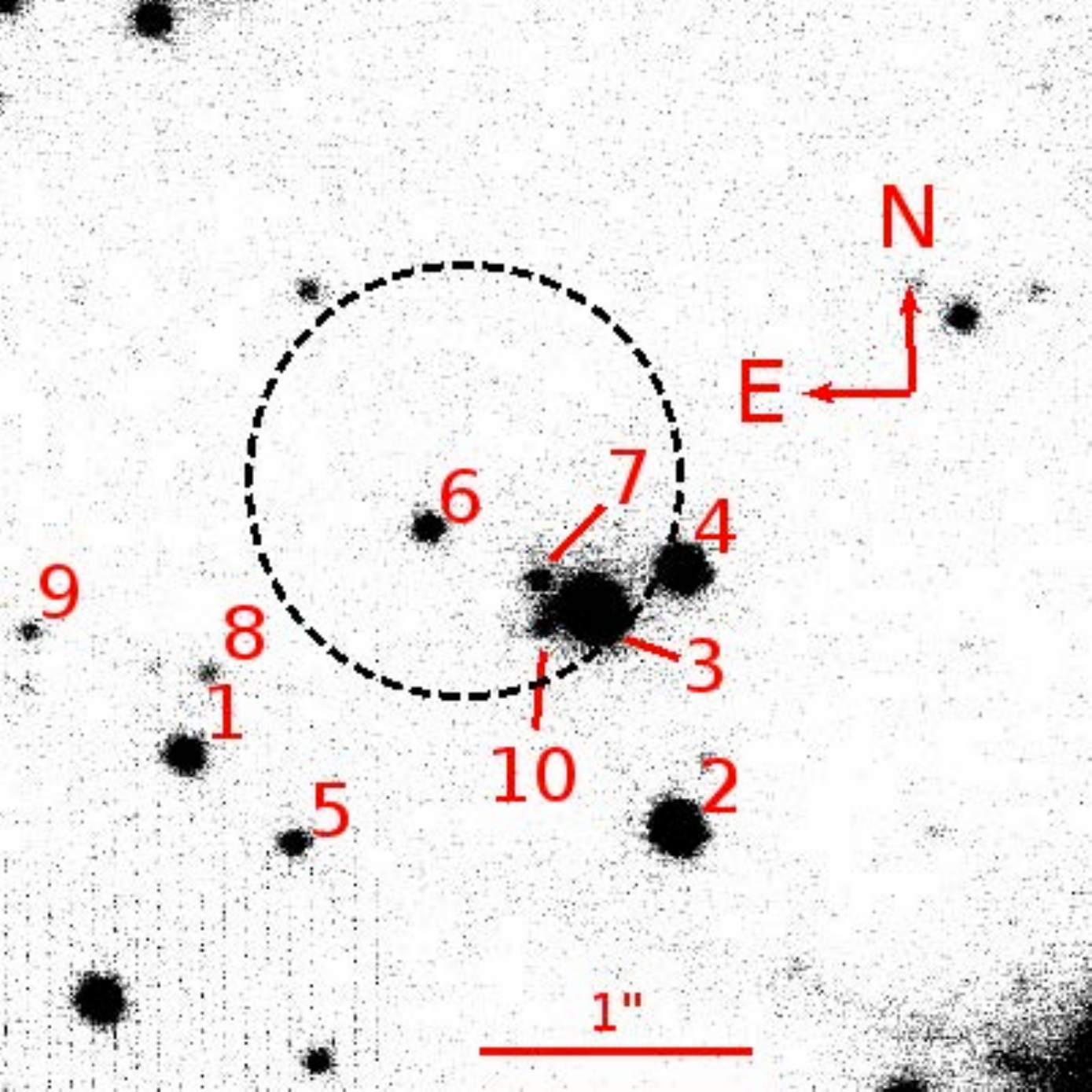}}
\caption{A $4\times4\,\mathrm{arcsec}$ cutout near SGR\,1900$+$14 from a $\mathrm{K_s}$ band LGS-AO supported observation from the NIRC2 camera. Stars are labelled as per \citet{testa2008}. The black circle is centered on the radio position of SGR\,1900$+$14 from \citet{frail1999} and encircles the 0.8$^{\prime\prime}$ radius, 99\%-confidence circle from \citet{testa2008} the positions from which are used for our absolute astrometry. Star 7 is the proposed counterpart of SGR\,1900$+$14 based on its variability.}
\label{fig:sgr1900_narrow}
\end{figure}

Figure~\ref{fig:sgr1900_narrow} shows a $4 \times 4$\,arcsec $\mathrm{K_s}$ band image from our LGS-AO observations with the NIRC2 camera around the X-ray position of SGR\,1900$+$14. The stars are labelled as per \citet{testa2008}. They obtained two $\mathrm{K_S}$ band AO observations of the same field around SGR\,1900$+$14 with VLT NACO instrument in March and July 2006. Star 7 was the only source inside the radio-position error circle (dashed circle in Figure~\ref{fig:sgr1900_narrow}) that showed a photometric variability. They detected a 3-$\sigma$ increase in the flux of star 7 and proposed it as the IR counterpart of SGR\,1900$+$14. We accept the counterpart proposed by \citet{testa2008}.

 In an attempt to gather additional evidence for the identification of the IR counterpart, we have measured $\mathrm{K_p}$ band photometric variability and $\mathrm{H}-\mathrm{K_p}$ colour for the stars in the field. These measurements are reported in Section~\ref{sec:sgr1900_results}. However, during this period, the X-ray counterpart did not show significant variablity. Hence the absence of NIR variation of the proposed counterpart does not provide any new insights. 

We report (in Section~\ref{sec:results}) that the proper motion of star 6 lies along the Galactic rotation curve, whereas the proper motion of star 7 is significantly different from those of galactic stars. This evidence strengthens the identification of star 7 as the IR counterpart of SGR\,1900$+$14.

\section{Observations and Analysis}
\label{sec:observations}
\subsection{Observations}

Starting in 2005 to the present time, we undertook a program for astrometric monitoring of magnetars with the 10-meter Keck 2 telescope using the Laser Guide Star Adaptive Optics~\citep[LGS-AO;][]{wizinowich2006,vandam2006} and the Near-Infrared Camera 2 (NIRC2). The log of our observations can be found in Tables~\ref{tab:sgr1806_observations} and \ref{tab:sgr1900_observations}.

 \begin{deluxetable}{cccc}  
\centering
\tablecolumns{4}
\tablecaption{Summary of observations of SGR\,1806$-$20.\label{tab:sgr1806_observations}}
\tablewidth{0pt}
\tabletypesize{\footnotesize}
\tablehead{   
  \colhead{Date \& MJD}&
  \colhead{Filt}      &
  \colhead{Cam}      &
  \colhead{Exp}  \\
  \colhead{(UTC MJD)}  &
  \colhead{}            &
  \colhead{}            &
  \colhead{($\mathrm{s}$)}}

\startdata
2005-03-04 53433.641 & $\mathrm{K_p}$      & N  & 1440       $\checkmark$ \\
2005-04-30 53490.511 & $\mathrm{K_p}$      & N  & 750         \\
2005-08-10 53592.366 & $\mathrm{K_p}$      & W  & 600         \\
2005-08-11 53593.344 & $\mathrm{K_p}$      & W  & 840         \\
2005-09-26 53639.258 & $\mathrm{K_p}$      & N  & 600       $\checkmark$  \\
2006-07-03 53919.403 & $\mathrm{K_p}$      & N  & 2820       \\
2006-08-17 53964.304 & $\mathrm{K_p}$      & N  & 1800        $\checkmark$ \\
2007-05-22 54242.487 & $\mathrm{K_p}$      & N  & 1020        \\
2007-06-11 54262.403 & $\mathrm{K_p}$      & N  & 2040        \\
2007-07-16 54297.345 & $\mathrm{K_p}$      & N  & 3000        \\ 
2007-08-06 54318.329 & $\mathrm{K_p}$      & N  & 2640       $\checkmark$  \\
2008-05-21 54607.468 & $\mathrm{K_p}$      & N  & 2460        \\
2008-06-29 54646.407 & $\mathrm{K_p}$      & N  & 3360         \\
2008-07-26 54673.342 & $\mathrm{K_p}$      & N  & 3180       $\checkmark$  \\
2010-06-18 55365.442 & $\mathrm{K_p}$      & W  & 80          \\

\enddata
\tablecomments{A $\checkmark$ in Column 4 marks~ the images used for astrometric measurements.}
\end{deluxetable}

\begin{deluxetable}{cccc}  
\centering
\tablecolumns{4}
\tablecaption{Summary of observations of SGR\,1900$+$14.\label{tab:sgr1900_observations}}
\tablewidth{0pt}
\tabletypesize{\footnotesize}
\tablehead{   
  \colhead{Date \& Time}&
  \colhead{Filt}      &
  \colhead{Cam}      &
  \colhead{Exp} \\
  \colhead{(UTC)}  &
  \colhead{}            &
  \colhead{}            &
  \colhead{($\mathrm{s}$)}}

\startdata
2005-04-30 53490.558 & $\mathrm{K_p}$      & N  &  1300    $\checkmark$ \\
2005-08-09 53591.434 & $\mathrm{K_p}$      & W  &  2400    \\
2005-08-10 53592.400 & $\mathrm{K_p}$      & W  &  300     \\
2005-09-26 53639.349 & $\mathrm{K_p}$      & W  &  720     \\
2006-07-03 53919.472 & $\mathrm{K_p}$      & N  &  1980    \\
2006-07-04 53920.511 & $\mathrm{K_p}$      & N  &  1920    \\
2006-08-17 53964.439 & $\mathrm{K_p}$      & N  &  1140    $\checkmark$ \\
2006-10-13 54021.242 & $\mathrm{K_p}$      & N  &  2220    \\
2007-05-22 54242.550 & $\mathrm{K_p}$      & N  &  1500    $\checkmark$ \\
2007-06-11 54262.553 & $\mathrm{K_p}$      & N  &  1260    $\checkmark$ \\
2007-06-11 54262.582 & $\mathrm{H}$        & N  &  660     \\
2007-08-06 54318.455 & $\mathrm{K_p}$      & N  &  1800    $\checkmark$ \\
2007-11-03 54407.229 & $\mathrm{K_p}$      & N  &  1260    $\checkmark$ \\
2008-05-21 54607.564 & $\mathrm{K_p}$      & N  &  1260    \\
2008-06-29 54646.471 & $\mathrm{K_p}$      & N  &  3660    \\
2008-07-26 54673.405 & $\mathrm{K_p}$      & N  &  2280    $\checkmark$ \\
2008-10-22 54761.227 & $\mathrm{K_p}$      & N  &  1680    $\checkmark$ \\
2009-04-06 54927.599 & $\mathrm{K_p}$      & N  &  1620    \\
2009-07-17 55029.340 & $\mathrm{K_p}$      & N  &  2100    $\checkmark$ \\
2009-08-04 55047.346 & $\mathrm{K_p}$      & N  &  2100    \\
2009-09-29 55103.226 & $\mathrm{K_p}$      & N  &  2340    \\
2010-06-18 55365.470 & $\mathrm{K_p}$      & N  &  2340    $\checkmark$ \\

\enddata
\tablecomments{A $\checkmark$ in Column 4 marks the images used for astrometric measurements.}
\end{deluxetable}

\subsubsection{NIRC2} The NIRC2 instrument has two modes: wide (W) and narrow (N) with a field-of-view (FoV) of $\approx 10\times 10\,$arcsecond and $\approx 40\times 40\,$arcsecond respectively. The corresponding pixel scales are 9.942\,milli-arcsecond per pixel and 39.768 milli-arcsecond per pixel. The wide field images were obtained to aid transferring the photometry and astrometry from the low resolution 2MASS images to the small FoV narrow camera NIRC2 images. The narrow field images were used for the astrometric measurements. Based on weather and faintness of each magnetar, multiple short ($\sim$20 s) exposures were chosen to avoid saturating the detector. The typical full width at half maximum (FWHM) achieved in these observations was $\approx 70\,$milli-arcsecond $\approx 7\,\mathrm{pix}$.

Each of the NIRC2 narrow camera images was inspected for quality control. Images in which the AO correction was poor were rejected. The shallow images with acceptable AO correction were rejected for astrometry due to the non-detection of the magnetar and/or lack of sufficient reference stars but were used to photometric calculate upper limits on the brightness. The images used in the final proper motion measurement are denoted by a $\checkmark$ in Column 4 of Tables~\ref{tab:sgr1806_observations} and \ref{tab:sgr1900_observations}.

\subsection{Data Analysis}
The images from the NIRC2 camera were reduced using the FITS analysis package \texttt{pyraf} in a standard manner by subtracting corresponding dark frames and flat-fielded using appropriate dome-flats. A sky fringe frame was made by combining dithered images of multiple targets with the bright stars masked. We used \texttt{SExtractor}~\citep{bertin1996_sextractor} for the preliminary detection and masking of stars. The fringe frame was subtracted after being scaled to the appropriate sky background level. Before coadding the frames, each frame was corrected for optical distortion using a distortion solution measured for NIRC2\footnote{See http://www2.keck.hawaii.edu/inst/nirc2/\\forReDoc/post\_observing/dewarp/}.

\subsubsection{PSF Fitting}

We used the \texttt{IDL} package \texttt{StarFinder}~\citep{diolaiti2000} to perform PSF estimation, fitting and subtraction. This code iteratively estimates a normalized PSF shape from user selected stars, while subtracting faint neighboring stars to minimize the contamination of the PSF estimate. \texttt{StarFinder} fits a constant PSF shape over the entire field of view (FoV). This assumption appears to work well for the NIRC2 narrow camera FoV. The uniformity of the PSF over the FoV also mitigates the errors from centroiding variable PSFs. 

AO PSFs differ from PSFs obtained from atmospheric seeing limited observations in two aspects: Firstly, because the AO correction decorrelates as a function of distance from the AO reference source (i.e. sodium laser beacon), the PSF varies radially across the field of view. Secondly, since AO correction cannot correct all of the wavefront errors caused by atmospheric turbulence, even on-axis, AO PSFs have a distinctive shape with a sharp diffraction-limited (FWHM $\sim \lambda/D_{tel}$) core and a wide (FWHM $\sim$ atmospheric seeing) shallow halo around it. For the Keck AO system, these components are $44\,$milli-arcsecond and $\sim 1\,\mathrm{arcsecond}$ respectively. The order of magnitude difference in size and brightness of the two components makes it challenging to accurately measure and subtract the PSF in the image. We describe how both these challenges are handled in the next paragraph.

To further reduce the effect of PSF variations, relative photometry and astrometry measurements were down-weighted farther away from the object under consideration. The details of the relative weighting are described in Section~\ref{sec:relative_astrometry}. The PSF model size was chosen to be 200 pixels ($1.95\,\mathrm{arcsecond}$) wide to encompass both the core and the halo of the PSF. The few brightest stars in each of the fields were used for estimating the halo contribution.

\subsubsection{Relative Astrometry}
\label{sec:relative_astrometry}
\citet{cameron2009} demonstrated a framework for high precision astrometry ($< 100\,\mathrm{\mu arcsecond}$) through an optimal estimation technique that availed the correlations in stellar position jitter. We use the same methodology with modifications for including the proper motions of the stars over multiple epochs and an appropriate weighting scheme.

The dominant source of astrometric error in the single epoch, short exposure images of \citet{cameron2009} was tip-tilt anisoplanatism. For our coadded long exposure images the tip-tilt anisoplanatism is averaged out. We constructed the covariance matrix theoretically using geometry of the field and a typical turbulence profile from Mauna Kea. The residual distortion of the NIRC2 distortion solution has a root-mean-square value of $1\,$milli-arcsecond. However the distortion residuals have higher values towards the edges\footnote{http://www2.keck.hawaii.edu/inst/nirc2/\\forReDoc/post\_observing/dewarp/}. To reduce the effect of residual distortion, especially in images with significant dithering, a separation-weighted measurement scheme (the $\theta$ term used below) was used to downweight stars far from the target.

To account for the proper motions of all the stars in the field, it was necessary to include the proper motion estimates in the framework and simultaneously estimate a least-squares fit for grid positions and proper motions. Given $N+1$ stars detected in the field, the measurement of the offset between the target star and each of the remaining stars results in a set of vectors at each of the $E$ epochs. 

The differential offsets between star 0 and the grid of $N$ reference stars at epoch $k$ is written as a single column vector,
$$\mathbf{d}_{0k} = [x_{01},\ldots, x_{0N}, y_{01},\ldots , y_{0N}]_k^T.$$
Here $x_{ij} = x_j-x_i$ is the distance between the $x$-coordinate of the $j^{\mathrm{th}}$ reference star and the $x$-coordinate of the $i^{\mathrm{th}}$ target star, and likewise for $y$. The goal of differential astrometry is to use $\mathbf{d}$ to determine the position of the target star with respect to the reference grid of stars at each epoch.

We use a linear combination of the elements of $\mathbf{d}$ with weights $\mathbf{W_i}$ to obtain the relative position of target star $i$ at epoch $k$, 
$$\mathbf{p}_{ik} = \mathbf{W}_i \mathbf{d}_{ik},$$
where, for example, the weight matrix for star 0, $\mathbf{W}_0$ is
$$ \mathbf{W}_0 =
\begin{bmatrix}
w_{xx,01} & \ldots & w_{xx,0N} & w_{xy,01} & \ldots & w_{xy,0N}\\
w_{yx,01} & \ldots & w_{yx,0N} & w_{yy,01} & \ldots & w_{yy,0N} \end{bmatrix}.$$

We calculated weights as follows: $w_{xx,ij}^{-1} = w_{yy,ij}^{-1} = \sigma^2_{ij}$. Here $\sigma^2_{ij} = \sigma^2_m +\sigma^2_{TJ}\theta_{ij}^2$, where $\sigma^2_{TJ}$ is the geometric mean of the parallel and perpendicular components of the tip-tilt jitter as defined in Equation 1 of \citet{cameron2009}; and $\theta_{ij}$ is the angular offset between the star $i$ and the star $j$. We have used the notation $w_{xy,0j}$ to denote the weighting of the offset from the target star ($i = 0$) to star indexed $j$ in the $y$ direction which is used to determine the $x$ component of the target's position, $\mathbf{p}$.

We assume a simple linear model for the stellar motion where $x = z_x + v_x t$. The differential offsets are thus a column vector,
$$
\mathbf{d}_0 = \begin{bmatrix} 
z_{x,1} + v_{x,1}t - (z_{x,0} + v_{x,0}t) \\
\vdots\\
z_{x,N} + v_{x,N}t - (z_{x,0} + v_{x,0}t) \\
z_{y,1} + v_{y,1}t - (z_{y,0} + v_{y,0}t) \\
\vdots\\
z_{y,N} + v_{y,N}t - (z_{y,0} + v_{y,0}t) \end{bmatrix}$$
and the unknown quantities are,

\begin{eqnarray}
\mathbf{b} = [z_{x,0}, &\ldots&, z_{x,N}, v_{x,0}, \ldots, v_{x,N}, \\
      \ldots, z_{y,0}, &\ldots&, z_{y,N}, v_{y,0}, \ldots , v_{y,N}]^T.
\end{eqnarray}

We solve for the variables $\mathbf{b}$ from the vector $\mathbf{d}$ given weights $\mathbf{W}$ in the least squares sense. For a given target, we use the same weights for all epochs. The overall $x$ and $y$ shifts of each image (i.e. the registration of the image) are fit as free parameters in this method. 

NIRC2 is mounted at the Nasmyth focus of the Keck II telescope. A field-rotator allows the observer to set the position angle of the instrument. Our default position angle was zero degrees (North is up and East is to the left on the detector). However, there are small errors in the setting of the field rotator as well as tracking errors.

To measure this, we chose the images obtained on May 22, 2007 as the reference image for both the targets. The reference images were chosen on the basis of good AO correction and image depth. We computed the rotation-angle and the plate-scale of the image at each epoch with respect to the reference image. We find that the rotation angle is within 0.5 degrees and the image scaling is within 0.1\% relative to those of the reference image. The stellar position grids were corrected for the measured rotation and plate-scale changes before measuring their proper motions.  

To understand the systematic effects caused by our choice of grid stars, we re-analyzed the centroiding data after randomly eliminating a selected number of stars from the reference grid. We compared the results to those obtained from our entire grid of stars. For example, by eliminating one randomly chosen star out of the 50 stars in the SGR\,1900$+$14 field, the proper motions of all other stars change by $\Delta(\mu_{\alpha},\mu_{\delta}) = (7.6 \pm 15.4, 17.1 \pm 13.7) \times 10^{-3}\,$milli-arcseconds\,yr$^{-1}$. This is much smaller than our statistical errors of $\sim1\,$milli-arcsecond\,yr$^{-1}$. Hence we conclude that the choice of our reference grid is robust and does not add significant errors to our measurements.

\subsubsection{Galactic Rotation}
\label{sec:galactic_rotation}
Since our relative astrometry framework calculates the proper motion of each object with respect to a grid of neighboring stars (i.e. with respect to the average motion of all other stars), it implicitly assumes that the net velocity of the field is zero. However, this is not true since the rotation of the Galaxy and the peculiar velocity of the Sun with respect to the local standard of rest (LSR) cause significant motions at the precision we seek. Our framework cannot measure the net velocity of the field without prior knowledge of the absolute motion of a few stars or equivalently, the absolute non-motion of an extra-galactic object in the field. 

To correct for this effect, we need to calculate the mean galactic proper motion of all the stars in the field along the line of sight given by Galactic longitude and latitude $(l,b)$. We modelled the differential rotation of the Galaxy and the local velocity of the Sun and calculated the effective proper motion of an object at a given position $(r,l,b)$ in the Milky Way, where $r$ is the distance away from the Sun. We made a model assuming the local velocity of the Sun to be $(U, V, W) = (10.0, 5.2, 7.2)\,\mathrm{km\,s^{-1}}$~\citep{dehnen1998} and that the Galaxy is rotating with a constant circular speed outside of $R_1 = 2\,\mathrm{kpc}$ of $220\,\mathrm{km\,s^{-1}}$, decreasing linearly inside of that $R_1$~\citep{binneytremaine2008}. We set the distance from the Sun to the center of the Galaxy to $R_0 = 8.0\,\mathrm{kpc}$~\citep{eisenhauer2003}. From the rotation curve, we calculate the Galactic proper motion $\vec{\mu}_{\mathrm{Gal}} = [\mu_{\alpha},\mu_{\delta}]_{\mathrm{Gal}}$ of objects at various distances ($ 1\,\mathrm{kpc} \leq r \leq 20\,\mathrm{kpc}$) in the direction $(l,b)$ of the magnetar that are moving with the Galactic flow. 

We estimate the number density of stars in the Milky Way using the model calculated by \citet{juric2008} using SDSS data. They fit a thin disk, thick disk and a halo to the SDSS data set and calculate the number density function based on their fit. Along the line of sight, the number of stars in our field at a distance $r$ from the Sun is proportional to $r^2\rho(R,Z)$, where $\rho(R,Z)$ is the number density of stars at the cylindrical coordinates $(R(r,l,b),Z(r,l,b))$ in the Milky Way. 

For a given field, we calculate the velocity of the field $\vec{\mu}_{\mathrm{Field}} = [\mu_{\alpha},\mu_{\delta}]_{\mathrm{Field}}$ as the integral of the proper motion weighted with the number density as described above. This gives,

$$\vec{\mu}_{\mathrm{Field}} = \frac{\int_{r_{min}}^{r_{max}}r^2\rho(R(r,l,b),Z(r,l,b))\times[\mu_{\alpha,\delta}(r,l,b)_{\mathrm{Gal}}]\mathrm{d}r}{\int_{r_{min}}^{r_{max}}r^2\rho(R(r,l,b),Z(r,l,b))\mathrm{d}r}. $$

\begin{deluxetable*}{llccc}
\centering
\tablecolumns{5} 
\tablecaption{Proper motions calculated from the Galactic rotation model as described in Section~\ref{sec:galactic_rotation}.\label{tab:gal_rot}}
\tablewidth{0pt}
\tabletypesize{\footnotesize}
\tablehead{
  \colhead{Object ID}                                &
  \colhead{Distance}                                 &
  \colhead{$(l,b)$}                 &
  \colhead{$\vec{\mu}_{\mathrm{Field}}$}               &
  \colhead{$\vec{\mu}_{\mathrm{Gal}}$}                 \\
  \colhead{}                                         &
  \colhead{}                                         &
  \colhead{}                               &
  \colhead{$[\mu_{\alpha},\mu_{\delta}]$}              &
  \colhead{$[\mu_{\alpha},\mu_{\delta}]$}             \\
  \colhead{}                                         &
  \colhead{(kpc)}                                    &
  \colhead{(deg)}                                    &
  \colhead{(milli-arcsecond\,yr$^{-1}$)}                               &               
  \colhead{(milli-arcsecond\,yr$^{-1}$)}                               }
\startdata
SGR\,1806$-$20 & $9 \pm 2$  & $(10.0,-0.2)$ &  $[3.0, 4.8]$ & $[4.2 \pm 0.9, 7.0 \pm 1.8]$ \\
SGR\,1900$+$14 & $12.5 \pm 1.7$ & $(43.0,+0.8)$ & $[2.7, 4.6]$ & $[2.7 \pm 0.2, 4.8 \pm 0.4]$ \\
\enddata
\end{deluxetable*}

Thus, the total proper motion of each object in the sky is $\vec{\mu}_{\mathrm{Sky},i} = \vec{\mu}_{\mathrm{R},i}+\vec{\mu}_{\mathrm{Field}}$. Table~\ref{tab:gal_rot} lists the calculated proper motion for the field and the Galactic proper motion for an object at the distance of the magnetar for both of the targets.

\subsubsection{Peculiar Motion}
We are interested in back-tracing the proper motion of the magnetar to identify its birthsite and estimate the time since it left the birthsite. The relevant motion for this measurement is the relative proper motion between the magnetar and its progenitor. A reasonable assumption is that the progenitor, likely a young massive star, was moving with the Galactic rotation curve. We define the peculiar motion of the magnetar as the difference between its total proper motion $\vec{\mu}_{\mathrm{Sky},i}$ and its expected Galactic proper motion $\vec{\mu}_{\mathrm{Gal}}$, i.e. $\vec{\mu}_{\mathrm{Sky},i} = \vec{\mu}_{\mathrm{Gal}} + \vec{\mu}_{\mathrm{Pec}}$. 

With this definition, the transverse velocity of the magnetar relative to its neighbourhood becomes $r |\vec{\mu}_{\mathrm{Pec}}|$ in a direction $\theta,\,\mathrm{s.t.}\,\tan(\theta) = (\mu_{\alpha}/\mu_{\delta})_{\mathrm{Pec}}$ East of North. 

\subsubsection{Photometry}
\label{sec:photometry}
\texttt{StarFinder} calculates flux estimates for stars in the field by scaling the normalized PSF model to best fit the image. We calculate the photometric zero-point (ZP) for each image by comparing the magnitudes of stars to the 2 Micron All Sky Survey (2MASS) Point Source Catalog~\citep{skrutskie2006_2MASS} and to published high-resolution studies of the fields which were anchored to the 2MASS catalog. The details of comparison stars for each field are given in Section~\ref{sec:results}.

\section{Results}
\label{sec:results}
\subsection{SGR\,1806$-$20}
\label{sec:sgr1806_results}
 We performed PSF fitting on the NIRC2 narrow camera images to identify 71 stars through 10 epochs. The positions of these 71 stars were used for relative astrometry.

\begin{figure}[htb]
\centering
\includegraphics[width=0.48\textwidth]{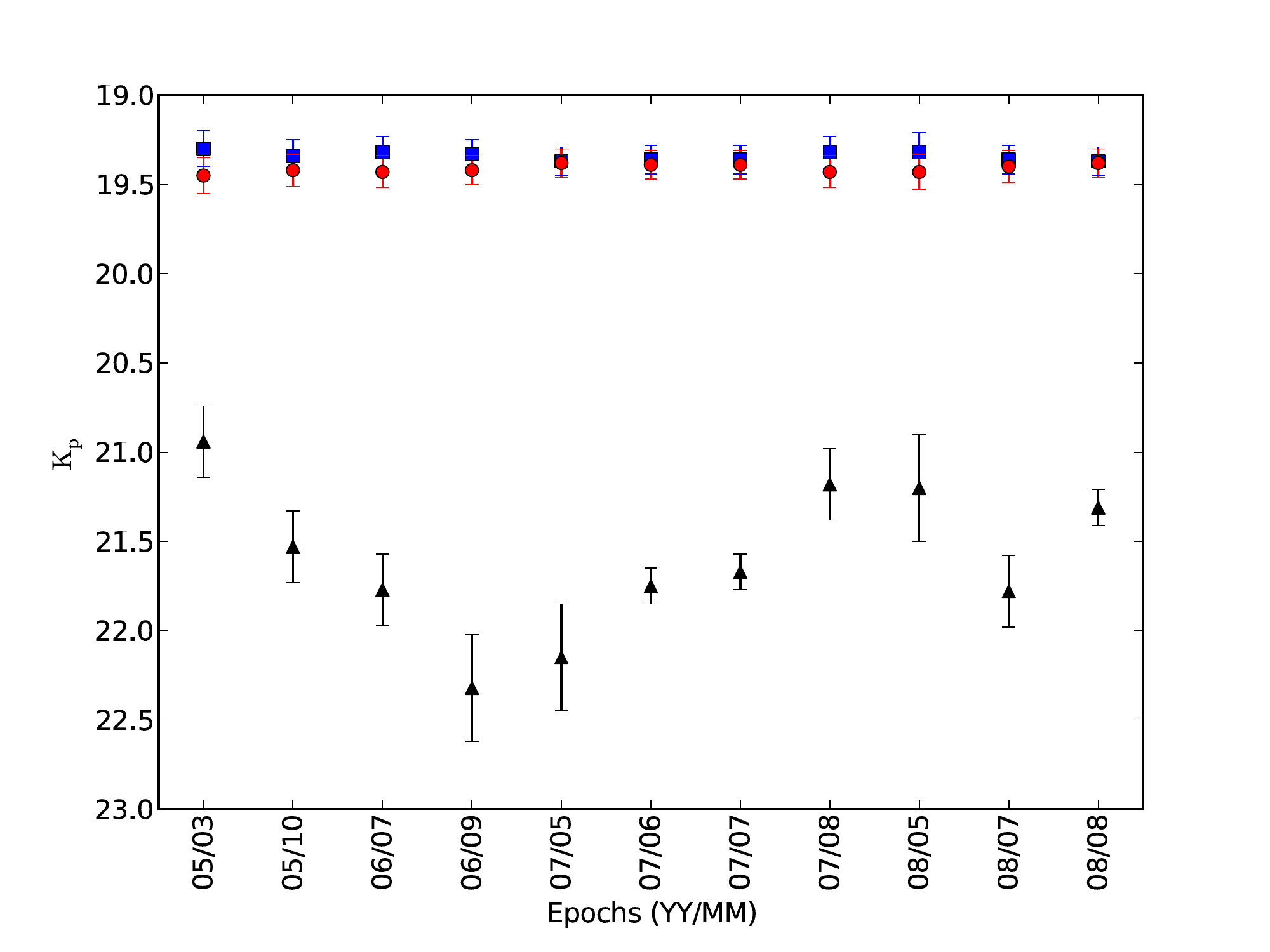}
\caption{$\mathrm{K_p}$ magnitudes of stars around SGR\,1806$-$20 measured over period of 3 years. The circles (red in the online version) correspond to star B and squares (blue in the online version) correspond to star C. The counterpart (star A) of SGR\,1806$-$20 is marked by black triangles. We note a clear variation over a factor of 3 in the brightness of star A.}
\label{fig:sgr1806_timeseries}
\end{figure}

 We performed relative photometry on the stars A, B and C in Figure~\ref{fig:sgr1806_IR_closeup}. The photometric zeropoints were measured by matching the magnitudes of stars B and C to the values measured by ~\citet{kosugi2005}. Figure~\ref{fig:sgr1806_timeseries} shows the measured magnitudes of the three stars. We observe a clear factor of 3 variation in the brightness of the IR counterpart of SGR\,1806$-$20, star A, thus securing the identification of the IR counterpart of SGR\,1806$-$20.

\subsubsection{Proper Motion}

\begin{figure}[htb]
\centering
\includegraphics[width=0.48\textwidth]{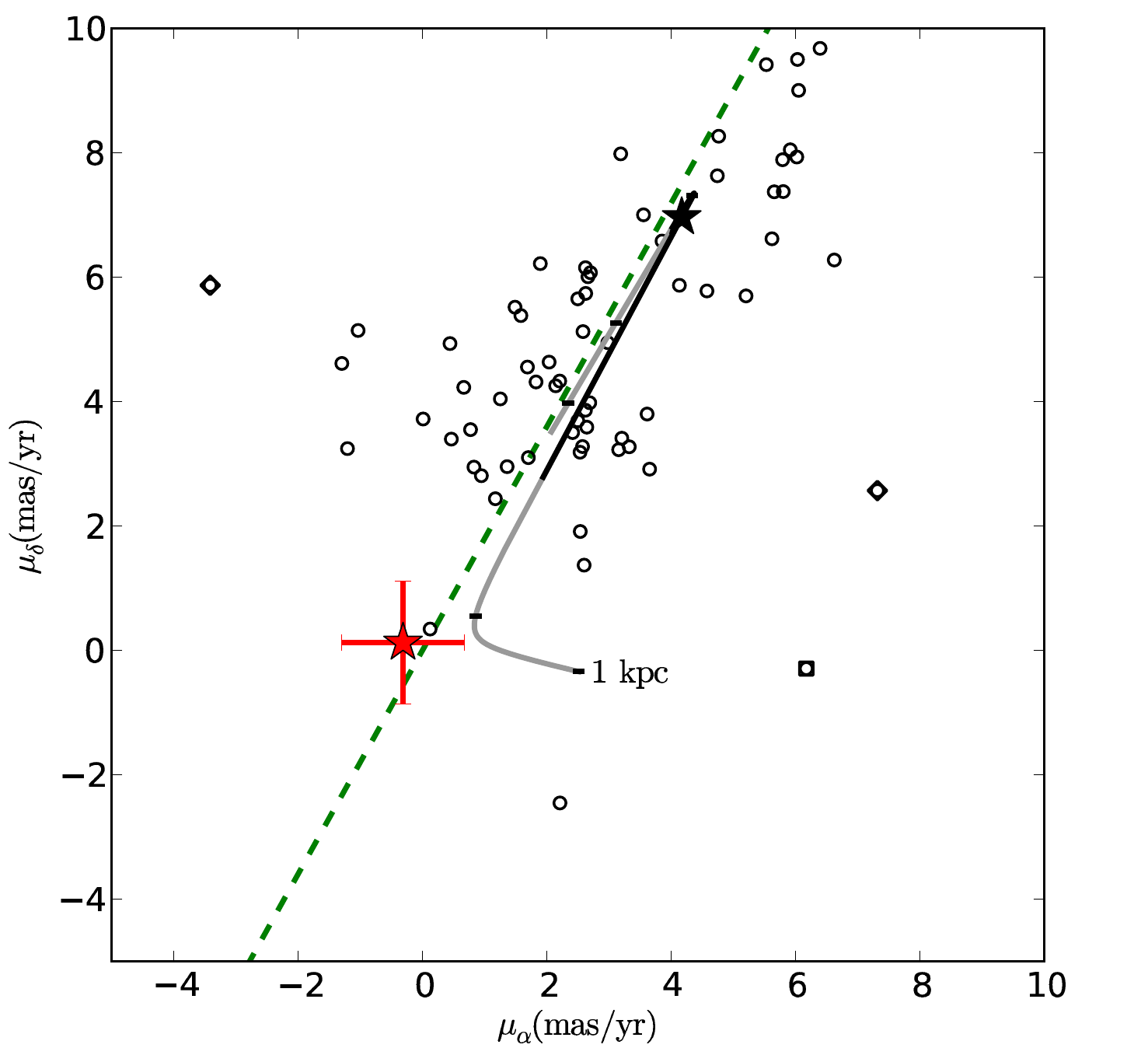}
\caption{The proper motion of 71 stars in the field of SGR\,1806$-$20 in the sky frame of reference. SGR\,1806$-$20 is marked by the star with error bars (colored red in the online version). The remaining stars have only their best-fit values (hollow black circles) after adding the bulk motion of the field ($\vec{\mu}_{\mathrm{Field}} = (3.0, 4.8)\,$milli-arcsecond\,yr$^{-1}$) (marked by a black `$+$'). The thick gray line represents the expected motion of stars from $1$ to $22.8\,\mathrm{kpc}$ along this line of sight, as per the Galactic rotation model presented in Section~\ref{sec:galactic_rotation}. Black dashes along the line denote positions 1, 5, 10, 15 and 20\,kpc away from the Sun. The section of the line representing objects at a distance of $9 \pm 2\,\mathrm{kpc}$ from the Sun is marked with a black star and black line to denote the possible motion of the progenitor of SGR\,1806$-$20.  The dashed diagonal line (green in the online version) is the locus of objects with $\mu_b = 0$, i.e. with zero proper motion along galactic latitude. Other high proper-motion objects, probably halo stars are marked by diamonds. The square marks the nominally high proper-motion object near the edge of the detector. However, this measurement may be corrupted by distortion residuals and hence is not considered any further.}
\label{fig:sgr1806_pm_disp}
\end{figure}

Figure~\ref{fig:sgr1806_pm_disp} shows the measured proper motions of the  stars in the SGR\,1806$-$20 field. The field velocity correction was calculated to be $(\mu_{\alpha},\mu_{\delta})_{\mathrm{Field}} = (3.0,4.8)\,$milli-arcsecond\,yr$^{-1}$. The proper motion of SGR\,1806$-$20 away from a putative progenitor in the galactic flow is $(\mu_{\alpha},\mu_{\delta}) = (-4.5\pm1.4,-6.9\pm2.0)\,$milli-arcsecond\,yr$^{-1}$. Assuming a distance of $9 \pm 2\,\mathrm{kpc}$, this corresponds to a linear velocity of $350 \pm 100\,\mathrm{km\,s^{-1}}$ with an angle of $213^{\circ} \pm 10^{\circ}$ East of North.

\begin{figure}[htb]
\centering
\fbox{\includegraphics[width=0.48\textwidth]{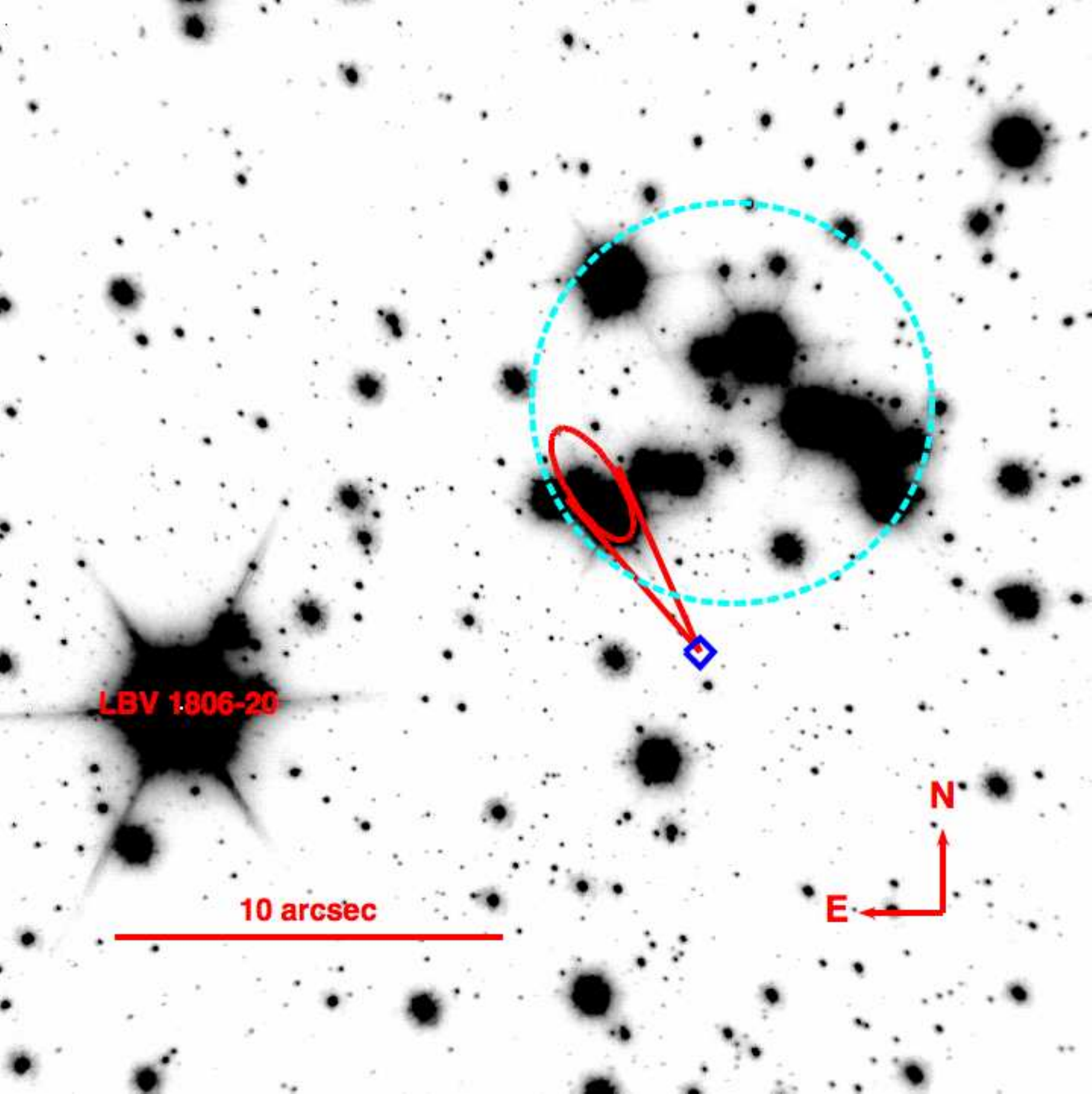}}
\caption{The position of SGR\,1806$-$20 (diamond, blue in the online version) traced back by 0.65\,kyr is marked by the ellipse (colored red in the online version). The size of the ellipse denotes the positional uncertainty corresponding to the uncertainty in the proper motion measurement. The solid lines (red in the online version) represent the 1-$\sigma$ limits on the angle of motion. The dashed circle (cyan in the online version) denotes the cluster of massive stars corresponding to the mid-IR source of \citet{fuchs1999}. The position of the luminous blue variable LBV 1806$-$20 is marked.}
\label{fig:sgr1806_motion_diagram}
\end{figure}

Figure~\ref{fig:sgr1806_motion_diagram} shows the direction of motion of SGR\,1806$-$20 with respect to its neighbors. Backtracing this space velocity would put the magnetar close to the cluster of massive stars about 650\,years ago.  

\subsubsection{Other High Proper-Motion Stars}
In Figure~\ref{fig:sgr1806_pm_disp}, we mark the high proper-motion objects with diamonds and squares. These stars deviate significantly from the dashed green line marking the locus of objects with $\mu_{b} = 0$, i.e. with zero proper motion along the galactic latitude. These are probably halo stars moving at a high speed through the Galactic disk.

\subsection{SGR\,1900$+$14}
\label{sec:sgr1900_results}
We observed SGR\,1900$+$14 at 13 epochs with an exposure time of about 1 hour at each observation. Using $\mathrm{K_p}$-band photometry and $\mathrm{H} - \mathrm{K_p}$ band color ( at a single epoch), we present variability and color measurements of SGR\,1900$+$14 and its surrounding stars. Our absolute astrometry is matched to positions as reported by \citet{testa2008} with an accuracy of 6 milli-arcsecond. They reported a 3-$\sigma$ astrometric uncertainty of 0.81$^{\prime\prime}$ which we adopt for comparison with the radio position for Figure~\ref{fig:sgr1900_narrow}.


In three images of the SGR\,1900$+$14 field that had excellent AO correction, we detected a faint source (labelled 10 in Figure~\ref{fig:sgr1900_narrow}) 0.2$^{\prime\prime}$ away from star 3. Source 10 is not detected by \citet{testa2008} as it was blended with star 3. However, we detected no variation in the combined brightness of star 3 and 10 in our data and the measurements from \citet{testa2008} within 0.07\,mag. Star 10 is a factor of $\sim40$ fainter than star 3. With this ratio, assuming no variation in the light from star 3, we can constrain the maximum variation in the brightness of star 10 to be 0.4\,mag as compared to the 0.48\,mag variation measured for star 7 and no variation for star 3 reported by \citet{testa2008}. Thus, we continue to accept star 7 as the IR counterpart of SGR\,1900$+$14.

\subsubsection{Variability}

\begin{figure}[htb]
\centering
\includegraphics[width=0.48\textwidth]{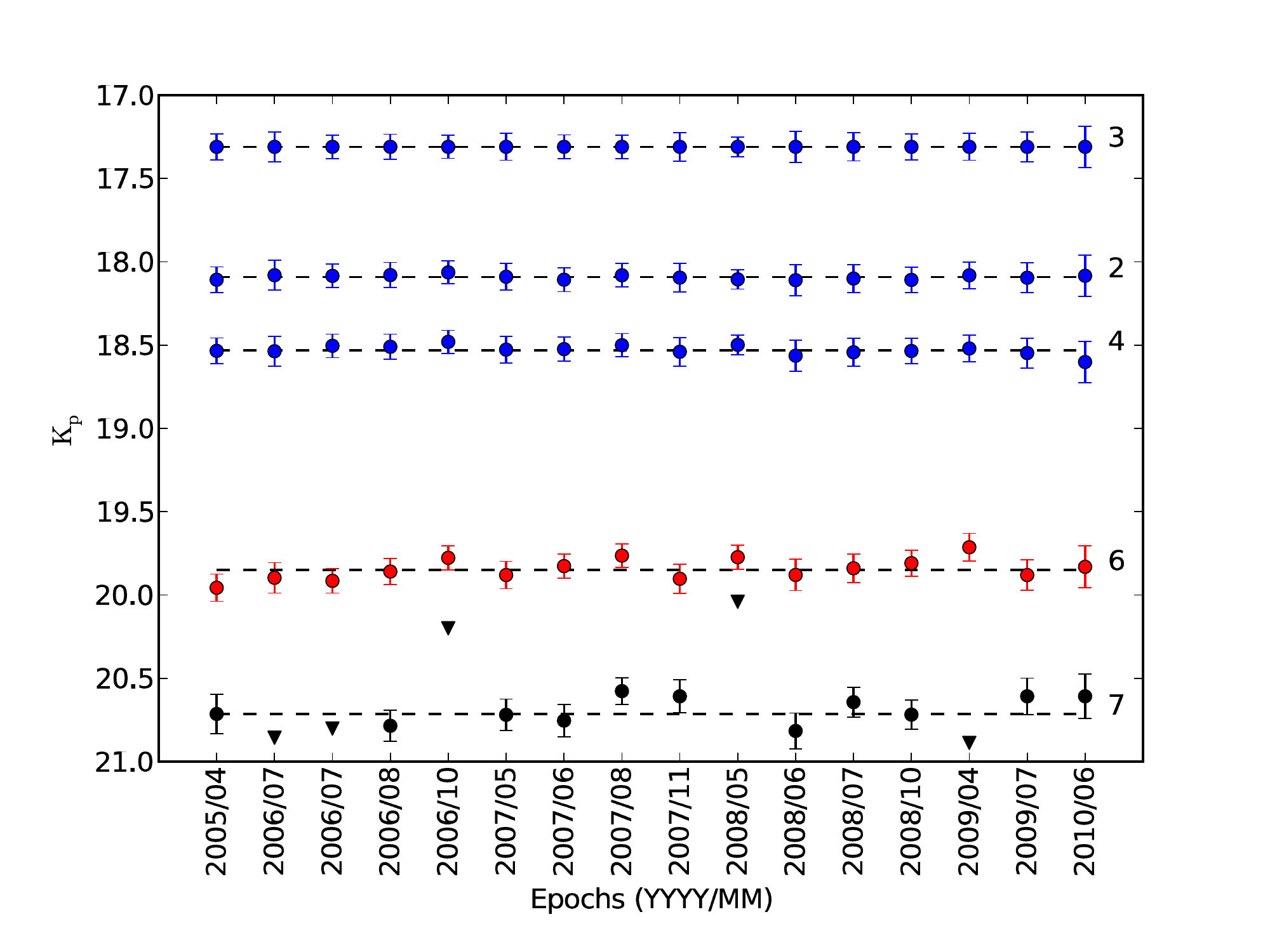}
\caption{Relative photometry light-curves of stars 2--7 (except 5) around SGR\,1900$+$14. To reduce the effect of PSF variations over the field, relative photometry was performed on nearby stars and the absolute calibration was performed by matching stars 2, 3 and 4 to their magnitudes as measured by~\citet{testa2008}. The inverted triangles mark  3-$\sigma$ upper limits for star 7 when it was not detected at the edge of star 3.}
\label{fig:sgr1900_timeseries}
\end{figure}

Figure~\ref{fig:sgr1900_timeseries} shows the photometry of stars 2--7 (except 5)\footnote{Star 5 is excepted from all further discussion since it is far away from the X-ray position error circle and does not affect any of the conclusions. Its identification in the middle of the numbering range is an unfortunate quirk of the numbering scheme that was implemented in previous literature.}. The median magnitude offsets of stars 2, 3 and 4 were used as relative ZP offsets and the absolute ZP offsets were calculated using $\mathrm{K_p}$ magnitudes as reported by \citet{testa2008}. The counterpart suggested by~\citet{testa2008}, star 7, was not detected at the edge of star 3 on epochs when the images were not sufficiently deep or the AO performance was not satisfactory. The non-detections were marked with the upper limit on the flux (black triangles). Including the upper limits on flux, star 7 shows slight variability but it is not conclusive. 

\begin{deluxetable}{lc}
\centering
\tablecolumns{2} 
\tablecaption{Persistent X-ray luminosity of SGR\,1900$+$14 in the 1-10\,keV band as reported by \citet{mereghetti2006} and \citet{israel2008}.\label{tab:sgr1900_xray_luminosity}}
\tablewidth{0pt}
\tabletypesize{\footnotesize}
\tablehead{
  \colhead{Interval}                         &
  \colhead{$F_X$}                       \\
  \colhead{(UTC Date)}                             &            
  \colhead{($10^{-12}\,\mathrm{erg\,cm^{-2}\,s^{-1}}$)}                             
}
\startdata
20 Sep 2005 - 22 Sep 2005 & $4.8 \pm 0.2$\tablenotemark{a} \\
25 Mar 2006 - 27 Mar 2006 & $4.6 \pm 0.8$\tablenotemark{b} \\ 
28 Mar 2006 - 28 Mar 2006 & $6.3 \pm 1.7$\tablenotemark{b} \\
01 Apr 2006 - 01 Apr 2006 & $5.5 \pm 0.4$\tablenotemark{a} \\
08 Apr 2006 - 10 Apr 2006 & $5.0 \pm 1.4$\tablenotemark{b} \\
11 Apr 2006 - 15 Apr 2006 & $5.0 \pm 0.7$\tablenotemark{b} \\
\enddata

\tablenotetext{a}{Absorbed 0.8-12\,keV flux from \citet{mereghetti2006}.}
\tablenotetext{b}{Unabsorbed 1-10\,keV flux from \citet{israel2008}.}
\end{deluxetable}

During our entire observation period from 2005 to 2010, the X-ray counterpart of SGR\,1900$+$14 showed burst activity in only one period from March to June 2006~\citep{israel2008}. Unfortunately, we have no IR observations between September 2005 and July 2006. Of these, the AO performance in July 2006 was not satisfactory leading to poor photometry and source confusion. As shown in Table~\ref{tab:sgr1900_xray_luminosity}, the persistent X-ray luminosity as measured by \citet{israel2008} and \citet{mereghetti2006} showed a slight increase in March 2006 and decreased to the pre-burst value by April 2006. Thus the lack of NIR variability is not surprising.

\subsubsection{Color Measurement}
\begin{figure}[htb]
\centering
\includegraphics[width=0.48\textwidth]{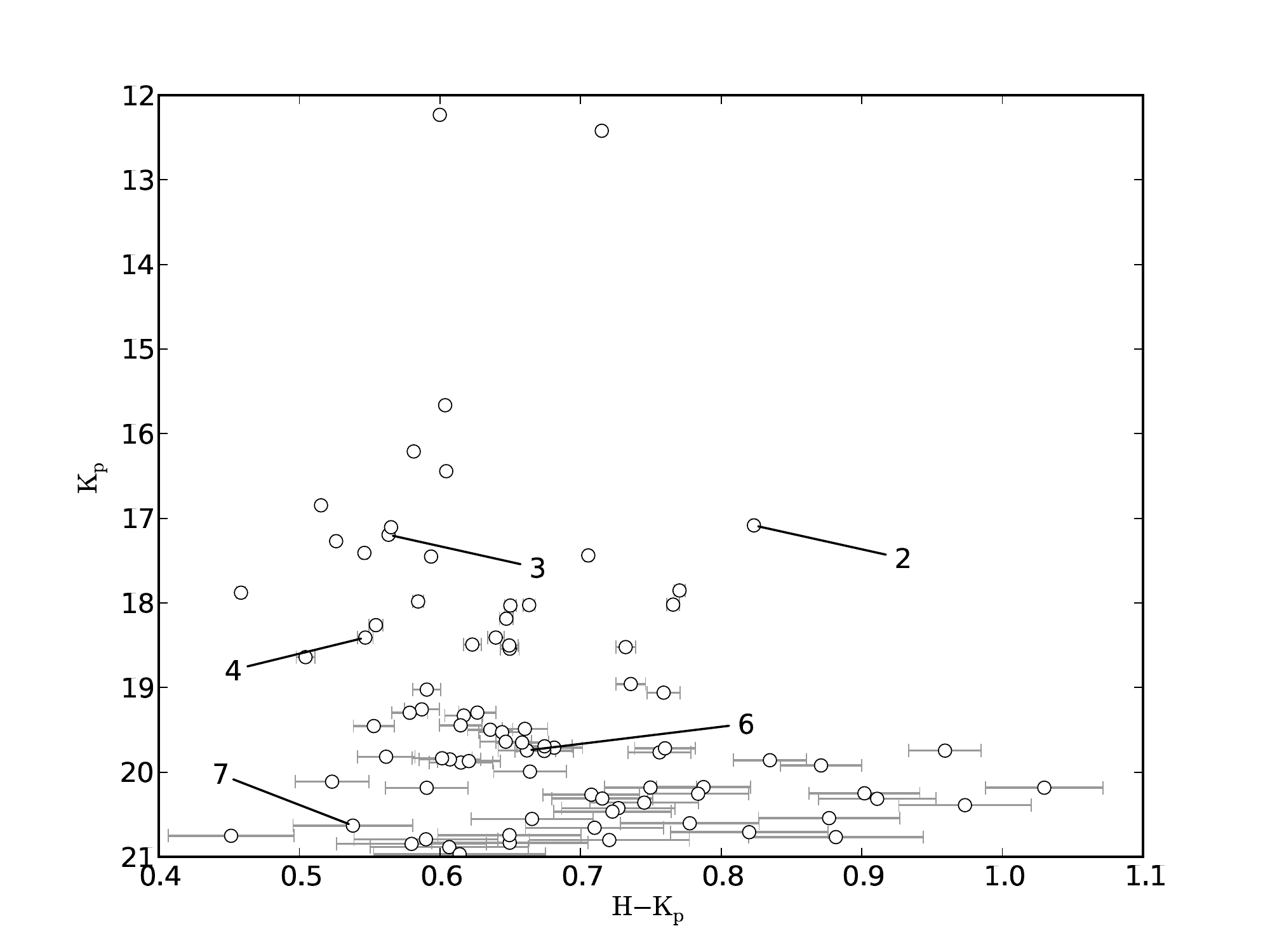}
\caption{$\mathrm{H}-\mathrm{K_p}$ color vs $\mathrm{K_p}$ magnitude diagram for 50 stars in the SGR\,1900$+$14 field. Stars 2--7 (except 5) are marked.The $\mathrm{H}$ band image zero-point has a systematic uncertainty of $\sim0.5\,\mathrm{mag}$ which would effectively only change the scale of the $x-$axis. }
\label{fig:sgr1900_cmd}
\end{figure}

During the June 11, 2007 observations, we obtained $\mathrm{K_p}$ and $\mathrm{H}$ band images of the field. These images were used to determine the colors of stars near SGR\,1900$+$14. No high-resolution $\mathrm{H}$ band photometry of this field has been performed previously, so we chose to use 2MASS measurements of bright stars to calculate the ZP offsets for the $\mathrm{H}$ band image. The problem with this implementation was that stars bright enough to be included in the 2MASS catalog were were saturated in the NIRC2 images which were intended to image the faint magnetar. We rely on the reconstruction of the saturated cores of bright stars by \texttt{StarFinder}. This increases the error in photometric measurement and hence in the ZP estimate. We estimate this systematic error in $\mathrm{H}$ band ZP to be $0.5\,\mathrm{mag}$. This systematic error changes the scaling on the $x$-axis of the color-magnitude diagram (Figure~\ref{fig:sgr1900_cmd}) and should not change the conclusion if the magnetar were to have a color distinctly different from other stars in the field.

\begin{deluxetable*}{llcccc}
\centering
\tablecolumns{6} 
\tablecaption{$\mathrm{H}$ and $\mathrm{K_p}$ band photometry for stars 2--7 (except 5) near SGR\,1900$+$14. The zero-point error in the photometry is 0.5$\,$mag for $\mathrm{H}$ band and 0.1$\,$mag for $\mathrm{K_p}$ band.\label{tab:sgr1900_H_kp}}
\tablewidth{0pt}
\tabletypesize{\footnotesize}
\tablehead{\colhead{Object ID}                         &
           \colhead{RA (J2000)}                        &
           \colhead{Dec (J2000)}                       &
           \colhead{$\mathrm{H}$ band}                 &
           \colhead{$\mathrm{K_p}$ band}               \\
           \colhead{}                                  &
           \colhead{(deg)}                             &
           \colhead{(deg)}                             &               
           \colhead{(mag)}                             &            
           \colhead{(mag)}                             }
\startdata
2 &    19$^{\mathrm{h}}$\,07$^{\mathrm{d}}$\,14.28$^{\mathrm{s}}$ & 9$^{\circ}$\,19$^{\prime}$\,18.84$^{\prime\prime}$ & $ 18.57 \pm 0.003  $ & $17.98 \pm 0.002$ \\
3 &    19$^{\mathrm{h}}$\,07$^{\mathrm{d}}$\,14.30$^{\mathrm{s}}$ & 9$^{\circ}$\,19$^{\prime}$\,19.63$^{\prime\prime}$ & $ 17.76 \pm 0.002  $ & $17.19 \pm 0.001$ \\
4 &    19$^{\mathrm{h}}$\,07$^{\mathrm{d}}$\,14.28$^{\mathrm{s}}$ & 9$^{\circ}$\,19$^{\prime}$\,19.78$^{\prime\prime}$ & $ 18.96 \pm 0.005  $ & $18.41 \pm 0.003$ \\
6 &    19$^{\mathrm{h}}$\,07$^{\mathrm{d}}$\,14.34$^{\mathrm{s}}$ & 9$^{\circ}$\,19$^{\prime}$\,19.92$^{\prime\prime}$ & $ 20.41 \pm 0.02   $ & $19.74 \pm 0.01 $ \\
7 &    19$^{\mathrm{h}}$\,07$^{\mathrm{d}}$\,14.31$^{\mathrm{s}}$ & 9$^{\circ}$\,19$^{\prime}$\,19.74$^{\prime\prime}$ & $ 21.17 \pm 0.04   $ & $20.63 \pm 0.02 $ \\
\enddata
\end{deluxetable*}

Figure~\ref{fig:sgr1900_cmd} shows an $\mathrm{H}-\mathrm{K_p}$ color vs. $\mathrm{K_p}$ magnitude diagram for the 50 stars in the field. Stars 2--7 are labeled. Neither star 6 nor star 7 have abnormal colors and neither is distinctive. There is no clear structure (for example, a main-sequence) in the color-magnitude diagram. This is probably due to the varied distances, ages and extinctions to the stars in this direction. Table~\ref{tab:sgr1900_H_kp} lists the $\mathrm{H}$ and $\mathrm{K_p}$ band magnitudes of stars 2--7 (except 5) as shown in Figure~\ref{fig:sgr1900_cmd}. Magnetars are not known to fall in a specific color band and our lack of understanding of the background physics prevents us from predicting the shape of the IR emission spectrum~\citep{testa2008}. We conclude that the lack of a distinctive color for any star near the location of SGR\,1900$+$14 is not significant.

\subsubsection{Proper Motion}

\begin{figure}[htb]
\centering
\includegraphics[width=0.48\textwidth]{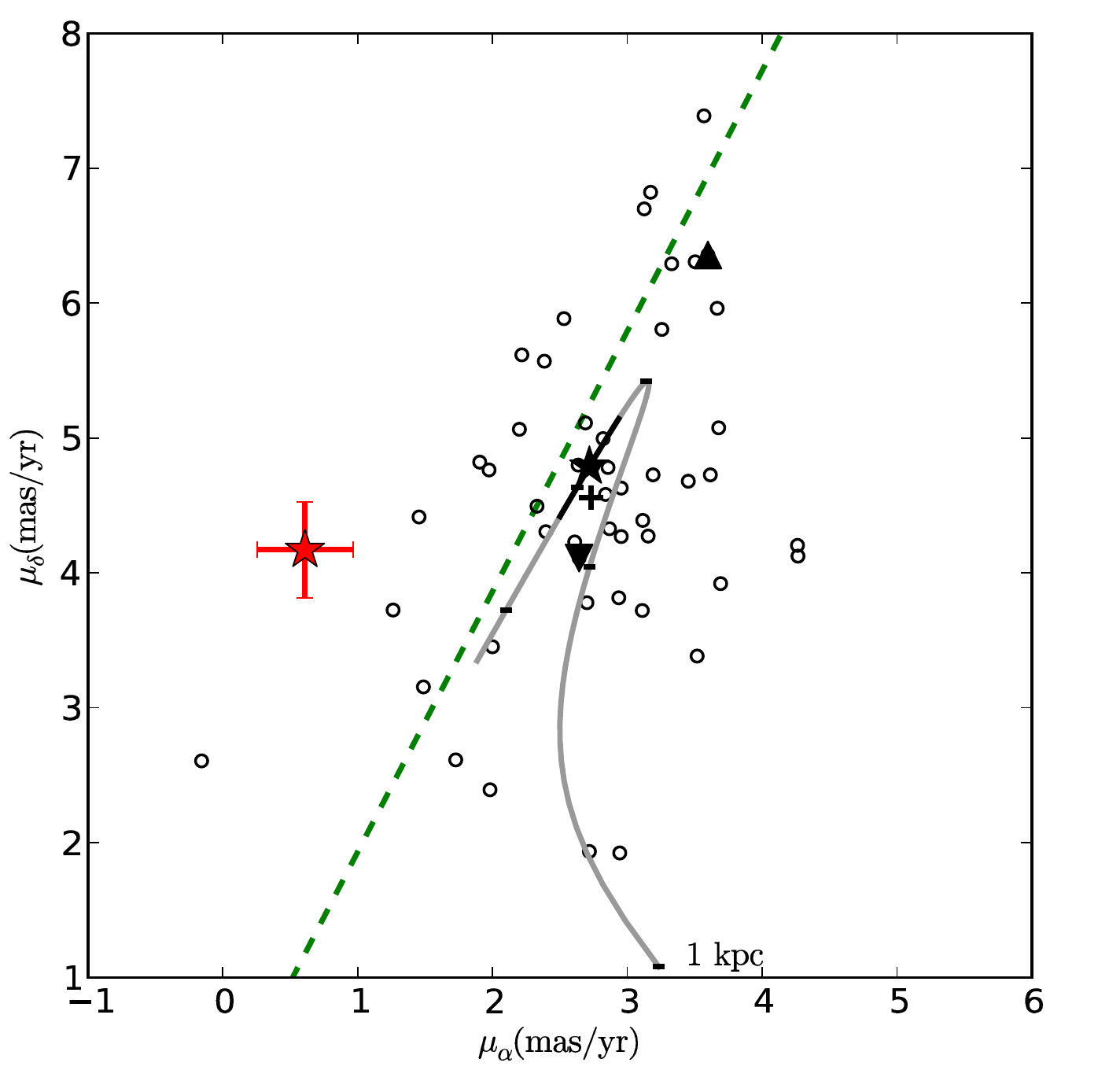}
\caption{The proper motion of 50 stars in the field of SGR\,1900$+$14 in the sky frame of reference. The putative counterpart of SGR\,1900$+$14 is marked by the star with error bars (colored red in the online version). The proper motions of star 6 (solid black triangle) and star 3 (inverted black triangle) seem to lie along the Galactic rotation curve. The remaining stars have only their best-fit values (hollow black circles) after adding the bulk motion of the field ($\vec{\mu}_{\mathrm{Field}} = (2.7, 4.6)\,$milli-arcsecond\,yr$^{-1}$) (marked by a black $+$). The thick gray line represents the expected motion of stars from $1$ to $19.8\,\mathrm{kpc}$ along this line of sight, as per the Galactic rotation model presented in Section~\ref{sec:galactic_rotation}. Black dashes along the line denote positions 1, 5, 10, 15 and 20\,kpc away from the Sun. The section of the line representing objects at a distance of $12.5 \pm 1.7\,\mathrm{kpc}$ from the Sun is marked with a black star and a black line to denote the possible motion of the progenitor of SGR\,1900$+$14.  The dashed diagonal line (green in the online version) is the locus of objects with $\mu_b = 0$, i.e. with zero proper motion along galactic latitude.}
\label{fig:sgr1900_pm_disp}
\end{figure}

Figure~\ref{fig:sgr1900_pm_disp} shows the measured proper motions of 50 stars in the neighborhood of SGR\,1900$+$14. The velocity offset, calculated from the galactic rotation, is $(\mu_{\alpha},\mu_{\delta})_{\mathrm{Field}} = (2.7,4.6)\,$milli-arcsecond\,yr$^{-1}$. For star 7, we calculate a proper motion of $(\mu_{\alpha},\mu_{\delta}) = (-2.1 \pm 0.4, 0.6 \pm 0.5)\,$milli-arcsecond\,yr$^{-1}$ away from a putative progenitor moving with the galactic flow. At a distance of $12.5\pm1.7\,\mathrm{kpc}$, this corresponds to a transverse space velocity of $130 \pm 30\,\mathrm{km\,s^{-1}}$. 

\begin{figure}[htb]
\centering
\fbox{\includegraphics[width=0.48\textwidth]{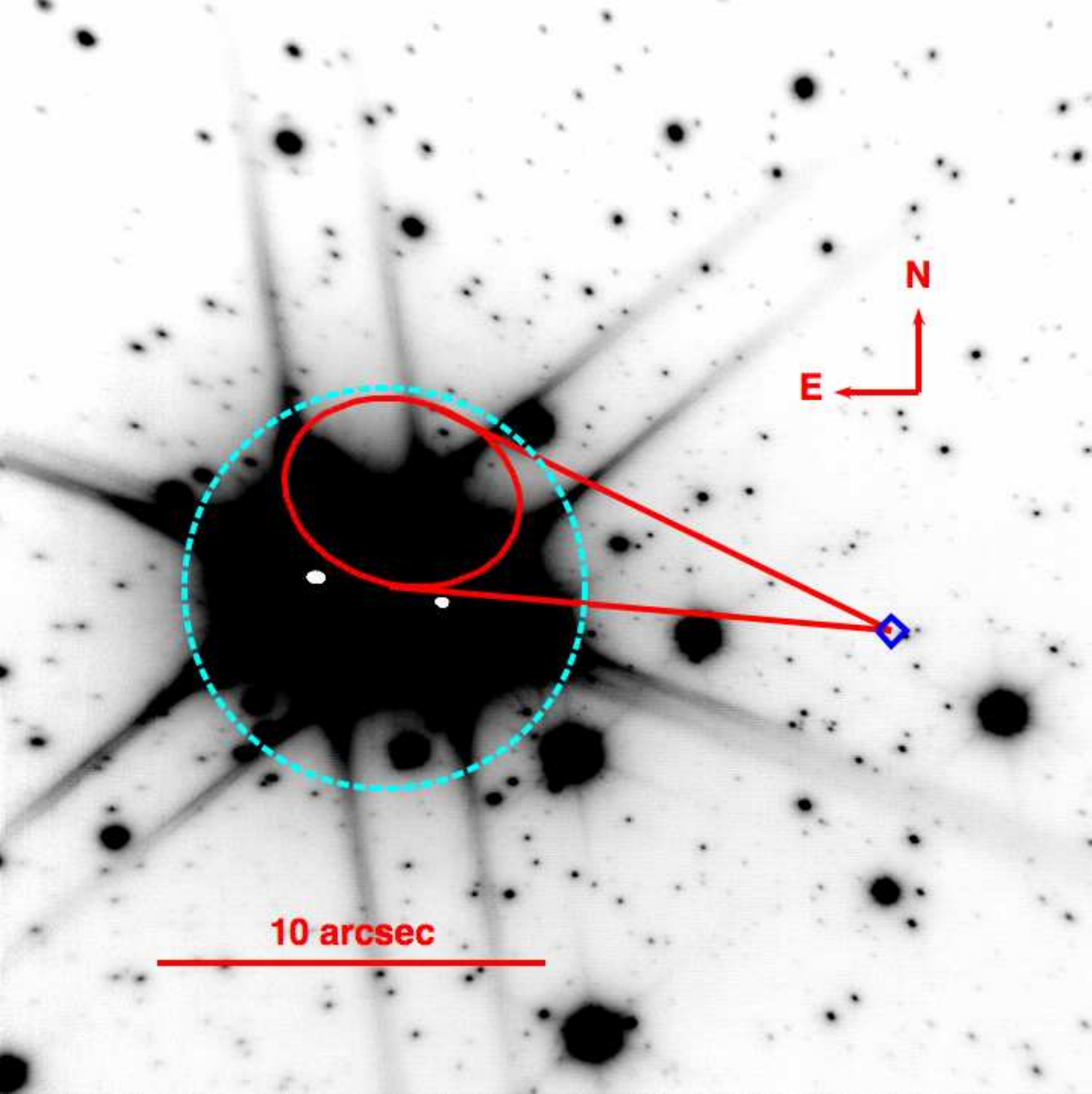}}
\caption{The position of the putative counterpart of SGR\,1900$+$14 (blue diamond) traced back by 6\,kyr is marked by the solid ellipse (red in the online version). The size of the ellipse denotes the positional uncertainty corresponding to the uncertainty in the proper motion measurement. The solid (red) lines represent the 1-$\sigma$ limits on the angle of motion. The dashed circle (cyan in the online version) denotes the cluster of massive stars~\citep{vrba2000}.}
\label{fig:sgr1900_motion_diagram}
\end{figure}

Figure~\ref{fig:sgr1900_motion_diagram} shows the direction of motion of SGR\,1900$+$14 with respect to its neighbors. Backtracing this space velocity would put the magnetar close to the cluster of massive stars about 6\,kyr ago. 

Star 6 and star 3 are the only two other sources detected inside the 3-$\sigma$ error circle around the radio position of SGR\,1900$+$14. Their velocities are marked by a black triangle (Star 6) and an inverted black triangle (Star 3) in Figure~\ref{fig:sgr1900_pm_disp}. Their velocities suggest that these are regular galactic stars moving in the plane of the galaxy (dashed green line). 

\begin{deluxetable}{lccc}
\centering
\tablecolumns{4} 
\tablecaption{Proper motions measured for stars 2--7 near SGR\,1900$+$14. The values have been corrected for the galactic rotation offsets. The transverse space velocities are calculated assuming a distance of $12.5\,\mathrm{kpc}$. 1-$\sigma$ error bars on $\vec{\mu}_{\mathrm{Pec}}$ are $(0.4,0.5)\,$milli-arcsecond\,yr$^{-1}$.\label{tab:sgr1900_propermotions}}
\tablewidth{0pt}
\tabletypesize{\footnotesize}
\tablehead{
  \colhead{Object}           &
  \colhead{$\vec{\mu}_{\mathrm{Pec}}$}   &
  \colhead{Velocity}                          &
  \colhead{Direction}                        \\
  \colhead{}           &
  \colhead{(milli-arcsecond\,yr$^{-1}$)}           &
  \colhead{(km\,s$^{-1}$)}                          &
  \colhead{E of N}                        
}
\startdata
2 & $(-0.11, -0.55) $  & $33\pm25$  & $ 191\pm143$ \\
3 & $(-0.08, -0.67) $  & $40\pm25$  &  \nodata  \\
4 & $(-0.74, -2.39) $  & $148\pm30$  & $ 197\pm10$ \\
6 & $(+0.88, +1.58) $  & $107\pm30$  & $ 30\pm12$ \\
7 & $(-2.11, -0.61) $  & $130\pm30$  & $ 254\pm10$ \\
\enddata
\end{deluxetable}

Table~\ref{tab:sgr1900_propermotions} gives the proper motions measured for each of the stars 2--7 along with their corresponding transverse space velocity assuming a distance of $12.5\,\mathrm{kpc}$.

\section{Discussion}
\label{sec:discussion}
Using LGS adaptive-optics supported near-IR observations, we have measured the proper motions of SGR\,1806$-$20 and SGR\,1900$+$14 to be $(\mu_{\alpha},\mu_{\delta})=(-4.5,-6.9)\pm(1.4,2.0)\,$milli-arcsecond\,yr$^{-1}$ and $(\mu_{\alpha},\mu_{\delta})=(-2.1,-0.6)\pm(0.4,0.5)\,$milli-arcsecond\,yr$^{-1}$ respectively. These correspond to a linear transverse velocity of $350\pm100\,\mathrm{km\,s^{-1}}$ and  $130\pm30\,\mathrm{km\,s^{-1}}$ respectively at the measured distances of their putative associations. Previously, using Very Long Baseline Interferometry (VLBI) at radio wavelengths, transverse linear velocities have been measured only for two magnetars: the AXP\,1E\,1810$-$197: $212\pm35~\mathrm{km\,s^{-1}}$~\citep{helfand2007} and the AXP\,PSR\,J1550$-$5418: $280\pm120\,\mathrm{km\,s^{-1}}$~\citep{deller2012}. The radio counterpart for AXP\,PSR\,J1622$-$4950 has been recently identified by \citet{levin2010} and would lead to an accurate proper motion measurement with VLBI. With the transverse velocity measurements for two AXPs and two SGRs in the $100-400\,\mathrm{km\,s^{-1}}$ range, it is highly unlikely that each of these objects has an extremely high radial velocity component. Hence we conclude that magnetars as a family do not possess the high space velocities ($\sim1000\,\mathrm{km\,s^{-1}}$) that were expected earlier (cf. Rothschild \&\ Lingenfelter 1996).   

Consider the space velocities of other families of neutron stars in contrast with magnetars. Canonical radio pulsars $(B\sim10^{11}\,\mathrm{G})$ have typical space velocities of $\sim200-300\,\mathrm{km\,s^{-1}}$~\citep{hobbs2005}. \citet{tetzlaff2010} traced the motions of 4 young, hot X-ray bright isolated neutron stars to associate them with progenitors and constrain their ages. They calculated the space velocities of these objects to be $\sim350\pm180\,\mathrm{km\,s^{-1}}$. There are a few fast moving pulsars such as PSR\,J1357$-$6429, which is a Vela-like radio pulsar has a transverse velocity of $1600-2000\,\mathrm{km\,s^{-1}}$ ~\citep{kirichenko2012}, but these seem to be outliers from the family. From these data, we observe that perhaps velocities are not a good discriminator of different groups of neutron stars and their origins.

\subsection{Association}
\label{sec:association}
Our measured proper motions provide very good evidence linking SGR\,1806$-$20 to the cluster of massive stars. The time required for SGR\,1806$-$20 to move from the cluster to its current position is $650\pm300\,\mathrm{yr}$. It may not be a surprise that one of the younger supernovae in our galaxy resulted from the magnetar. However, SGR\,1806$-$20 lies in the galactic plane behind dust clouds which create very high extinction in the visible wavelengths. Hence, the supernova associated with the magnetar may not have been visible to the naked eye. For SGR\,1900$+$14, we rule out any association with the supernova remnant G\,42.8$+$0.6 and confirm that this magnetar is associated with the star cluster. The time to trace the magnetar back to the cluster is  $6\pm1.8\,\mathrm{kyr}$.

The turn-off masses for the clusters with which the magnetars are associated allow us to place lower limits on the progenitor masses of these magnetars. Currently, progenitor mass estimates exist for three of the magnetars:\\
SGR\,1806$-$20: $48^{+20}_{-8}\,\mathrm{M_{\odot}}$~\citep{bibby2008},\\
CXO\,J1647$-$455: $>40\,\mathrm{M_{\odot}}$~\citep{muno2006,ritchie2010} and\\ 
SGR\,1900$+$14: $17 \pm 2\,\mathrm{M_{\odot}}$~\citep{davies2009}.\\

We note that only the two youngest SGRs have a star cluster in their vicinity. The lack of a star cluster in the vicinity of the older SGRs (despite ages of 4 to 10 kyr) suggests that it is not essential that SGRs should be associated with star clusters. Furthermore, the inferred progenitor masses of SGR 1900+14 does not compel us to believe that SGRs arise from massive stars. We conclude that binarity likely has a bigger role in forming SGRs.

\subsection{Braking Index}
\label{sec:braking_index}
If the association of the SGRs with the star clusters is taken for granted, we can constrain the braking index of the magnetars. The braking index $n$ is calculated from the following implicit equation: $$n = 1 + \frac{P}{T\dot{P}}(1-(P_0/P)^{(n-1)}).$$ Here, $T$ is the kinematic age of the magnetar (time taken to move from cluster to present position) and $P_0$ is the spin period at birth.

The instantaneous $\dot{P}$ is known to vary by a factor of three to four corresponding to large variations of braking torque on the magnetar~\citep{woods2002, woods2007}. We use the X-ray timing measurements from \citet{kouveliotou1998,mereghetti2005b,woods2007,marsden1999,woods2002,woods2003,mereghetti2006,nakagawa2009} to calculate an average $\dot{P}$ of $49\,\times\,10^{-11}\,\mathrm{s\,s^{-1}}$ for SGR\,1806$-$20 and $17\,\times\,10^{-11}\,\mathrm{s\,s^{-1}}$ for SGR\,1900$+$14 from 1996 to 2006.

Assuming $P_0/P \ll 1$, we estimate $n$ to be $1.76^{+0.65}_{-0.24}$ for SGR\,1806$-$20 and $1.16^{+0.04}_{-0.07}$ for SGR\,1900$+$14. This is significantly smaller than the canonical value of $n=3$ for the magnetic dipole spindown mechanism for pulsars. Low braking indices have been discussed in the context of twisted magnetospheres~\citep[eg. ][]{thompson2002} and particle wind spindown~\citep[e.g. ][]{tong2012}. However, the large variations in $\dot{P}$ over tens of years implies that these measurements cannot be taken at face value.

\acknowledgements{We would like to thank M. van Kerkwijk and C. Thompson for their critical comments and extensive discussions. The data presented herein were obtained at the W.M. Keck Observatory, which is operated as a scientific partnership among the California Institute of Technology, the University of California and the National Aeronautics and Space Administration. The Observatory was made possible by the generous financial support of the W.M. Keck Foundation.}

Facilities: \facility{Keck:II(NIRC2)}, \facility{Keck:II(LGS AO)}





\bibliographystyle{apj}
\bibliography{paper}

\end{document}